\documentclass[12pt]{spieman}  
\usepackage{amsmath,amsfonts,amssymb}
\usepackage{graphicx}
\usepackage{setspace}
\usepackage{tocloft}
\usepackage{todonotes}
\usepackage{array}
\usepackage{lineno}
\usepackage{textcomp}
\usepackage{hyperref}
\usepackage{multirow}
\usepackage{makecell}

\cftpagenumbersoff{figure}
\cftpagenumbersoff{table} 

\title{Cryogenic optical beam steering for superconducting device calibration}

\author[1,2,3]{K.~Stifter}
\author[1,2,3,4,5]{H.~Magoon}
\author[1,6,7]{A.J.~Anderson}
\author[1]{D.J.~Temples}
\author[1,2,3]{N.A.~Kurinsky}
\author[1]{C.~Stoughton}
\author[8,1]{I.~Hernandez}
\author[2,3,4]{A.~Nuñez}
\author[8,1]{K.~Anyang}
\author[1]{R.~Linehan}
\author[1,6,7]{M.R.~Young}
\author[9]{P.~Barry}
\author[1,10]{D.~Baxter}
\author[1]{D.~Bowring}
\author[1]{G.~Cancelo}
\author[1]{A.~Chou}
\author[6,7,1]{K.R.~Dibert}
\author[10,1]{E.~Figueroa-Feliciano}
\author[1]{L.~Hsu}
\author[8,1]{R.~Khatiwada}
\author[1]{S.D.~Mork}
\author[1]{L.~Stefanazzi}
\author[2,3,4]{N.~Tabassum}
\author[1]{S.~Uemura}
\author[11]{B.A.~Young}

\affil[1]{Fermi National Accelerator Laboratory, Batavia, IL 60510, USA}
\affil[2]{SLAC National Accelerator Laboratory, Menlo Park, CA 94025, USA}
\affil[3]{Kavli Institute for Particle Astrophysics and Cosmology, Stanford University, Stanford, CA 94305, USA}
\affil[4]{Department of Physics, Stanford University, Stanford, CA 94305, USA}
\affil[5]{Department of Physics and Astronomy, Tufts University, Medford, MA 02155, USA}
\affil[6]{Kavli Institute for Cosmological Physics, University of Chicago, IL 60637, USA}
\affil[7]{Department of Astronomy and Astrophysics, University of Chicago, IL 60637, USA}
\affil[8]{Department of Physics, Illinois Institute of Technology, Chicago, IL 60616, USA}
\affil[9]{School of Physics and Astronomy, Cardiff University, Cardiff, CF24 3YB UK}
\affil[10]{Department of Physics and Astronomy, Northwestern University, Evanston, IL 60208, USA}
\affil[11]{Department of Physics, Santa Clara University, Santa Clara, CA 95053, USA}

\begin{document} 
\maketitle

\begin{abstract}
We have developed a calibration system based on a micro-electro-mechanical system (MEMS) mirror that is capable of delivering an optical beam over a wavelength range of 180 -- 2000\,nm (0.62 -- 6.89\,eV) in a sub-Kelvin environment. 
This portable, integrated system can steer the beam over a $\sim$3\,cm\,$\times$\,3\,cm area on the surface of any sensor with a precision of $\sim$100\,$\mu$m, enabling characterization of device response as a function of position. 
This fills a critical need in the landscape of calibration tools for sub-Kelvin devices, including those used for dark matter detection and quantum computing.
These communities have a shared goal of understanding the impact of ionizing radiation on device performance, which can be pursued with our system.
This paper describes the design of the first-generation calibration system and the results from successfully testing its performance at room temperature and 20\,mK.
\end{abstract}


\begin{spacing}{1}

\section{Introduction}   

Astrophysical evidence indicates that approximately 85\% of all matter in the universe is ``dark"~\cite{Kolb:1990vq,2020A&A...641A...6P}, yet scientists have yet to uncover the fundamental nature of this dark matter beyond its gravitational interactions. 
So far, a host of experiments aiming to detect dark matter scattering have set strong bounds on particle-like dark matter models with masses above 1\,GeV/c$^2$ in the absence of a positive dark matter signal~\cite{Cooley:2022ufh, LZ, PICO:2019vsc, PDG}. 
This is prompting new searches to look over a broader range in dark matter masses~\cite{SnowmassLowThresholdReport, SENSEI, DAMIC-M}.  
Of particular interest are searches for particle-like dark matter with masses well below 1\,GeV/c$^2$. 
Development of sensors with the capability to detect scatters from this ``sub-GeV dark matter" requires thresholds at the eV scale and below~\cite{SnowmassLowThresholdReport}. 
Phonon- and other quasiparticle-sensitive sub-Kelvin superconducting devices such as transition-edge sensors (TESs)~\cite{Ren:2020gaq}, microwave kinetic inductance detectors (MKIDs)~\cite{temples2024performance,Cardani:2021iff,Cruciani:2022mbb}, superconducting nanowire single-photon detectors (SNSPDs)~\cite{Luskin:2023ksc}, and superconducting qubits~\cite{linehan2024estimating,Fink:2023tvb} are advantageous in this regime. 
In-situ characterization of device response to both backgrounds and to a putative dark matter signal is essential to validating the discovery potential of these detectors.

Independently, quantum computing has made great strides in the last decade through the improvement of coherence times of superconducting qubits, but a significant challenge remains in the reduction of error-inducing environmental noise. 
It has been shown that MeV-scale ionizing radiation produces correlated qubit errors~\cite{Wilen,Harrington:2024iqm} and shortened coherence times~\cite{Vepsalainen:2020trd,McEwen}. 
Though attempts to mitigate the sensitivity to these effects show great promise~\cite{McEwen:2024nrv,Iaia_2022}, a significantly improved understanding of the underlying mechanisms that cause superconducting qubit errors is needed~\cite{Martinis:2020fxb}.

Ionizing radiation is a primary background source for both the dark matter search and quantum computing communities, which share the need for a compact, robust, and sensitive calibration setup capable of producing $\sim$eV-scale energy deposits at a variety of specific locations across the surface of a device with a precision approaching the scale of features in the qubit architecture. 
Furthermore, superconducting qubits, TESs, KIDs, and SNSPDs typically operate at temperatures well below 1\,K, necessitating a calibration scheme that works at these temperatures or is able to penetrate a cryostat.
Historically, dark matter detectors have been calibrated through various schemes including radioactive sources, direct exposure to an LED, photons coupled through a fixed fiber, and filtered blackbodies~\cite{Baxter:2022dkm,temples2024performance,10.1063/5.0022533}, but each technique has key drawbacks.
Specifically, they either have energies too high to be useful for some modern applications, create energy deposits with rather limited position control or flexibility, are difficult to reproduce, and/or require coupling to a cryostat in ways that often introduce additional background events. 
These limitations motivated our development of an in-situ, steerable, collimated beam of low-energy photons that could be mounted at the mixing chamber stage (MXC) of a dilution refrigerator (DR).

Optical beam steering in a cryogenic environment is challenging due to both the dissipation of heat generated when the steering system is operated, as well as properties of the motion control mechanism that may cause it to stop functioning at low temperatures.
Existing optical steering methods include micro-electro-mechanical-system (MEMS) mirrors~\cite{Moffatt:2016kok,10.1063/1.4939753,10.1063/1.5131171,10.1063/5.0038392}, various piezoelectric systems~\cite{Moffatt:2016kok,2023APS..APRT12005O,PhysRevLett.119.223602}, galvo-scanners coupled to confocal microscopes~\cite{10.1117/12.2595780}, and cryogenic actuators~\cite{Benevides:2023ldf}. 
These systems all have trade-offs between system size, cryogenic functionality, heat production from power dissipation, scanning speed, range and precision, steering reproducibility, and introduction of parasitic backgrounds from light leakage.

In this paper, we introduce a full calibration system based on the MEMS mirror steering technology for use with any photon-sensitive device from room temperature down to 20\,mK and below.
The specifications for this design are shown in Table~\ref{tab:scanningMethods}, where we also list relative advantages and disadvantages of other optical beam steering systems compared to ours. 
In particular, our system achieves a steering range that allows for full coverage of cm$^{2}$- and in$^{2}$-scale devices in a way that limits parasitic backgrounds, while allowing for sub-Kelvin functionality and a small beam spot size.
Section~\ref{S:design} of the paper details the design of our full calibration system, and Section~\ref{S:results} describes the results from warm and cold operation of the system. 
This is followed by a discussion of follow-up work in Section~\ref{S:discussion} and conclusions in Section~\ref{S:conclusion}.

\begin{table}[t]
\begin{center}
    \caption{A selection of specific implementations of several optical steering methods. The specifications of the system presented in this paper are given in the top line, followed by some relative advantages and disadvantages of other systems for calibration of sub-Kelvin devices.}
       \begin{tabular}{|c|ccc|cc|}
            \hline
            \textbf{\makecell{Steering\\system}} & \multicolumn{1}{c|}{\textbf{\makecell{Steering\\range}}} & \multicolumn{1}{c|}{\textbf{Wavelength}} & \textbf{Spot size} & \multicolumn{1}{c|}{\textbf{Min. Temp.}} & \textbf{Power} \\
            \hline
            \textbf{\makecell{MEMS mirror\\(this work)}} & \multicolumn{1}{c|}{$\sim$3\,cm\,$\times$\,3\,cm} & \multicolumn{1}{c|}{180--2000\,nm} & $\sim$100\,\textmu m & \multicolumn{1}{c|}{\textless{}20\,mK} & $\sim$1\,\textmu W \\ 
            \hline \hline
            \textbf{} & \multicolumn{3}{c|}{\textbf{Advantages}} & \multicolumn{2}{c|}{\textbf{Disadvantages}} \\ 
            \hline
            \textbf{\makecell{MEMS mirror\\(prior work)~\cite{Moffatt:2016kok,10.1063/1.4939753,10.1063/1.5131171,10.1063/5.0038392}}} & \multicolumn{3}{c|}{60\,\textmu m spot size} & \multicolumn{2}{c|}{\makecell{\textgreater600\,mK operation,\\high light leakage}} \\ 
            \hline
            \textbf{\makecell{Piezoelectric\\systems~\cite{Moffatt:2016kok}}} & \multicolumn{3}{c|}{} & \multicolumn{2}{c|}{\makecell{High heat production,\\hysteresis, reduced\\deflection at low temp.}} \\ 
            \hline
            \textbf{\makecell{Attocube\\nanopositioner~\cite{2023APS..APRT12005O}}} & \multicolumn{3}{c|}{\makecell{42\,\textmu m spot size, target imaging,\\focal length control}} & \multicolumn{2}{c|}{\makecell{0.005\,cm$^{2}$ steering range,\\1\,K operation}} \\ 
            \hline
            \textbf{\makecell{Attocube\\nanopositioner~\cite{PhysRevLett.119.223602}}} & \multicolumn{3}{c|}{\makecell{3\,\textmu m resolution, target imaging,\\focal length control, free-space delivery}} & \multicolumn{2}{c|}{\makecell{0.3\,cm$^{2}$ steering range,\\high light leakage,\\high heat production}} \\ 
            \hline
            \textbf{\makecell{Scanning\\ confocal\\microscope~\cite{10.1117/12.2595780}}} & \multicolumn{3}{c|}{\makecell{\textless{}1\,\textmu m spot size, target imaging,\\ focal length control, free-space delivery}} & \multicolumn{2}{c|}{\makecell{0.5\,cm$^{2}$ steering range,\\ high light leakage,\\high heat production}} \\ 
            \hline
            \textbf{\makecell{Cryogenic\\actuator~\cite{Benevides:2023ldf}}} & \multicolumn{3}{c|}{\makecell{47\,\textmu m spot size, target imaging}} & \multicolumn{2}{c|}{0.003\,cm$^{2}$ steering range} \\ 
            \hline
        \end{tabular}
        \label{tab:scanningMethods}
    \end{center}
\end{table}
\section{Design }\label{S:design}

\begin{figure}[t]
    \centering
    \includegraphics[width=1\textwidth]{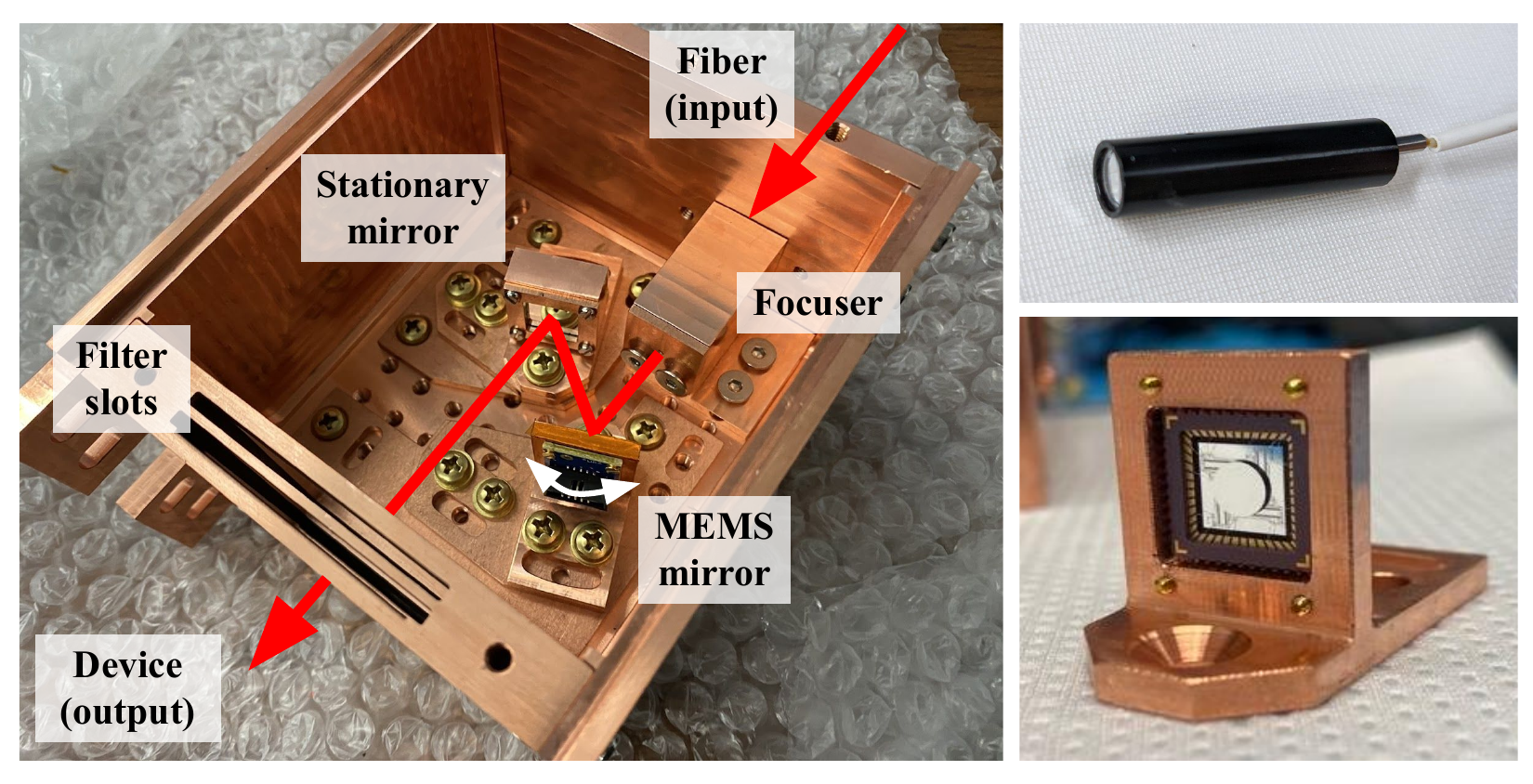}
    \caption{ 
    \textbf{Left:} Internal view of MEMS-based calibration system for sub-Kelvin operation. The high-conductivity copper enclosure, shown without its lid, has dimensions of 11\,$\times$\,10\,$\times$\,7\,cm. It includes translation plates for fine adjustment of optical components and light-tight ridges for reduction of backgrounds. Light, shown with a red arrow, emerges from a fiber-coupled refractive focuser \textbf{(top right)} and is reflected off of a steerable MEMS mirror \textbf{(bottom right)} from Mirrorcle Technologies, Inc. The MEMS mirror is used to accurately steer the incident beam onto a flat stationary mirror, which then directs the beam out of the enclosure via an IR filter window onto the surface of the device under test. 
    }
    \label{fig:enclosure}
\end{figure}

To meet the calibration needs of modern sub-Kelvin sensors, we designed and tested a calibration system capable of steering a collimated beam of light across the surface of any low-temperature sensor. 
The compact system is shown in Figure~\ref{fig:enclosure}. 
The deflection angle of the beam is dynamically controlled by a MEMS mirror located on the base temperature stage of a DR and enclosed in a 11\,$\times$\,10\,$\times$\,7\,cm housing machined from high-conductivity copper. 
Communication with the mirror control mechanism is performed using a simple computer interface at room temperature.
Although previous work was done using these mirrors at $\sim$1\,K~\cite{Moffatt:2016kok,10.1063/1.4939753,10.1063/1.5131171,10.1063/5.0038392}, we needed a system compatible with testing devices operating at $\sim$10\,mK, which required significant redesign. 
In our new system configuration, shown on the left in Figure~\ref{fig:enclosure}, an assembly of focusing and beam steering components are compactly housed inside a copper enclosure designed for light tightness, ease of handling, and minimal footprint within the DR.  The system was designed to be used with any fiber-coupled light source and a wide variety of test devices.  

The MEMS mirrors used (shown on the bottom right in Figure~\ref{fig:enclosure}) are commercial components purchased from Mirrorcle Technologies, Inc. The deflection angle of these mirrors can be controlled about two axes, allowing for steering across two dimensions, referred to as ``X" and ``Y" in this paper.  The circular face of the mirror is aluminized, allowing for operation over optical wavelengths and out into the THz regime~\cite{naftaly10.1364/AO.50.003201}. The MEMS mirrors are capacitive devices controlled by four voltages (two for each axis of rotation) in the range of 0\,-\,180\,V, with the angle of deflection set by the voltage difference across each axis.  This capacitive control provides the advantage that under ideal conditions, no power is dissipated while the mirrors are stationary, and heat is dissipated only when the mirror is moving.   In practice, some dissipation occurred even with the mirror stationary, as discussed in Section~\ref{S:results}. To modify the mirrors for cryogenic functionality, we worked with Mirrorcle Technologies to deposit aluminum over the doped silicon control lines. Without this modification, the control lines would disconnect at cryogenic temperatures due to carrier freeze out, and the lines would become high impedance, preventing the electronics from actuating the mirror. This also ensures that the control lines are superconducting below 1\,K and will minimize parasitic heating on the mirror substrate during charge or discharge of the MEMS.

In our experiments, the control voltages are generated by a commercial driver box purchased from Mirrorcle Technologies, capable of producing 180~V of positive bias on each of the four DC lines.  The DC control voltages are routed to the MEMS via cryogenic ribbon cables.  With this control system, an (X, Y) pair of voltage differences reproducibly translates to a precise location on a detector surface. In this paper, we describe our steering range in terms of this voltage coordinate system.  We define voltage coordinates such that the mirror's rest position is assigned [0,0], and voltage differences within the normalized range [-1,1] can be applied to both the X and Y direction of motion.

\begin{figure}[t]
    \centering
    \includegraphics[width=0.95\textwidth]{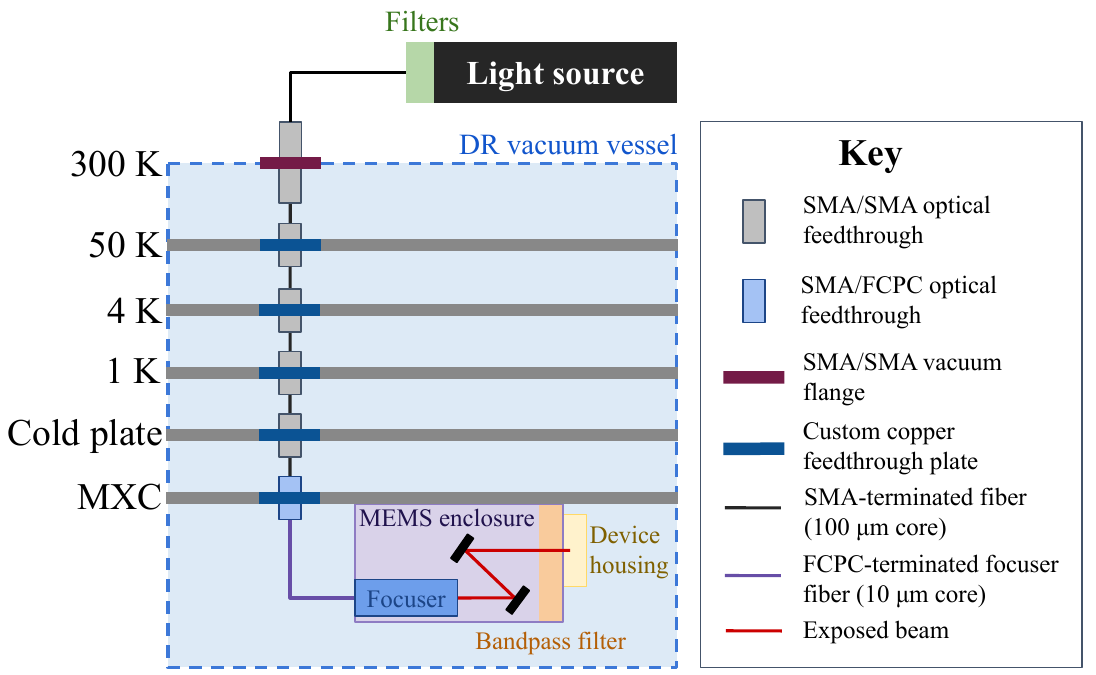}
    \caption{The optical path used to route photons from a room-temperature light source, into the DR cryostat, and to the MEMS enclosure mounted on the mixing chamber stage (MXC). The optical input line is broken at each DR stage to ensure thermalization of the fibers. Filtering after the warm light source and before the device housing reduces IR leakage and attenuates the total photon flux.}
    \label{fig:optical path}
\end{figure}

The setup was designed to be utilized with a variety of light sources, including both pulsed and continuous-wave beams over a wide range of wavelengths.  Optical fibers are used to route the beam from a room-temperature light source to the cryogenic steering unit.  A simple drawing of the optical path is shown in Fig. \ref{fig:optical path}. 
 The beam is first collimated from a light source at room temperature into an optical fiber.  This fiber passes through a KF40-to-SMA optical vacuum feedthrough into the cryostat.  The fibers are then routed through custom KF40 copper plates that thermalize the fiber at each cryostat stage.  From here, the fibers are connected to a refractive focuser, and routed internally in the steering system. We use 100\,\textmu m-core diameter multi-mode fibers that were purchased commercially from Accu-Glass Products, Inc. and transmit in the UV and visible wavelengths.  

The steering unit housing is designed as a copper box with a user-configurable optical breadboard-style base-plate.  Mounted to the breadboard is the refractive focuser, the MEMS mirror, and a stationary mirror.  Light enters the enclosure via the optical fiber and is directed through the refractive focuser.  This focuser is an OZ Optics DTS0060 commercial product.  It is coupled to a 10\,\textmu m-core optical fiber.  Light emerges from the focuser in the form of an exposed beam incident on the face of the MEMS mirror.  The beam is subsequently reflected off of a secondary, stationary mirror and directed through a filtered output window and onto the surface of the device under test.  The secondary mirror functions to extend the pathlength of the deflected beam, thus increasing steering area without requiring a larger physical enclosure for the system. The entire copper enclosure is mounted to the base temperature ($\sim 10$\,mK) stage of a dilution refrigerator.  The steering unit and test device mount is designed to be entirely light-tight.  Copper ridges at all the joints prevent stray photons from entering unit and interfering with the operations of the device under test. All of the electrical lines needed for MEMS control enter the light-tight box via a single MDM25 feedthrough.

To ensure adequate initial alignment of the laser system, the individual structural mounts for both the MEMS mirror and the stationary mirror were designed to allow for three degrees of translational freedom. While setting up the calibration unit, one can use these stages to optimize beam alignment and focus on the surface of the device under test.  The refractive focuser is mounted to a fixed position in the enclosure to allow for secure coupling to external fibers.  An external contact plate is used to apply clamping force to the component, maintaining its position and thermal contact with the copper housing.  A front plate with a small cutout is set in front of the focuser to further control optical alignment. 

\begin{figure}[t]
    \centering
    \includegraphics[width=0.9\textwidth]{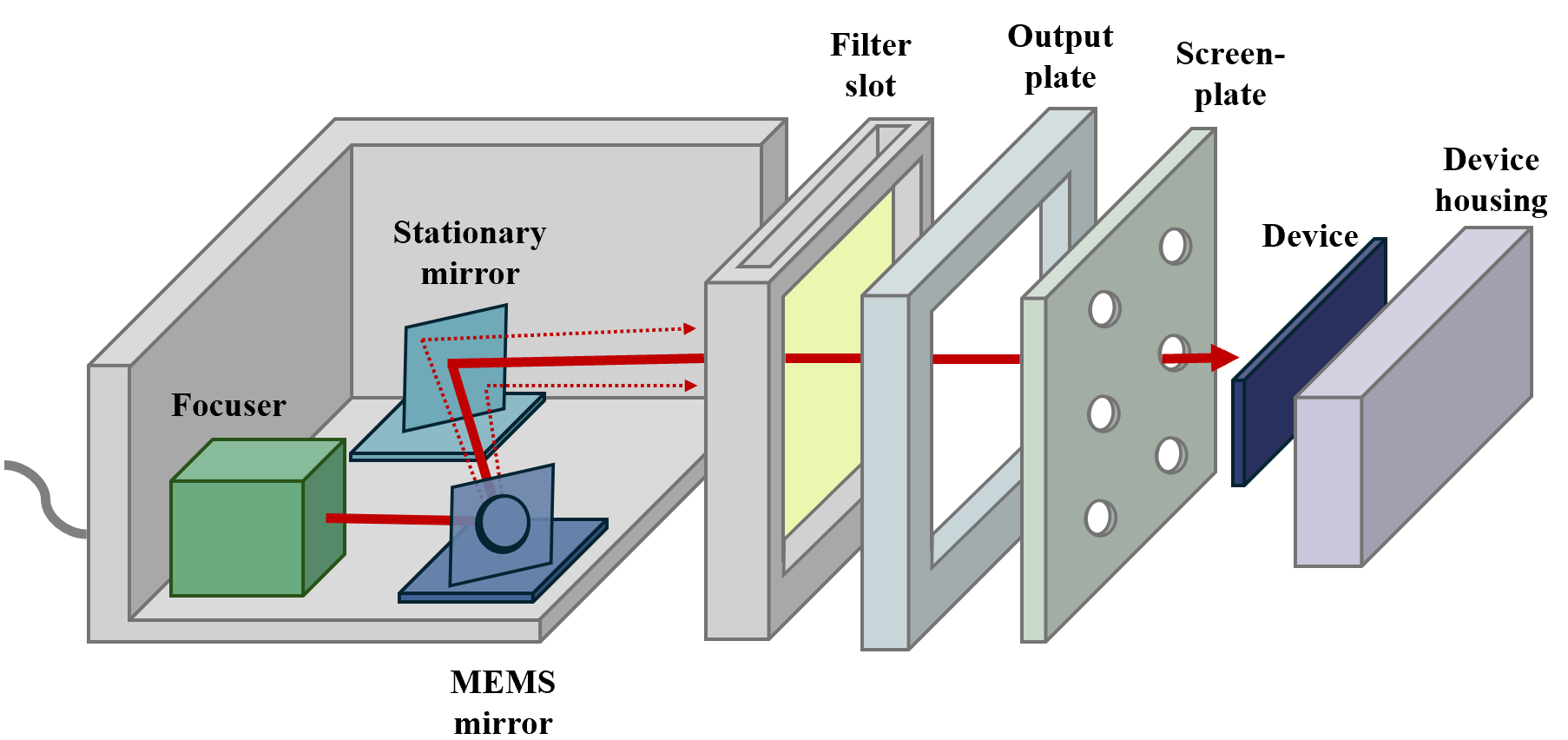}
    \caption{Photon path through the MEMS enclosure to the surface of a device under test. Light enters the MEMS enclosure via fiber-coupled refractive focuser and emerges as a beam in free space. The beam is reflected off a MEMS mirror which can sweep the beam across a 2D plane, and is then reflected off a stationary mirror to redirect it out of the enclosure. The beam passes through an IR filter and an experiment-specific ``output plate'' that is thermally anchored to the MXC of the DR before impinging on the surface of the device under test, which is located at the focal length of the focuser. A optional screen-plate can be used to restrict device illumination area.} 
    \label{fig:output}
\end{figure}

The laser beam exits the steering unit via a filter window as shown in Figure~\ref{fig:output}.  This window contains two slots for standard 2x2” square infrared (IR) filters, which are used to filter potential blackbody radiation of the optical fiber at higher temperature stages.  Thermal contact between the filters and the steering unit is maintained via the installation of a copper ``output plate".  The output plate is mounted to the MEMS enclosure using beryllium-copper spring washers.  In this configuration, pressure is applied uniformly to the square IR filters, enabling appropriate thermal contact between the filter and the housing.  To ensure sufficient thermalization of the device under test, copper brackets are used to connect the output plate directly to the mixing chamber at 10\,mK.  The thickness of the output plate can be designed such that the surface of the device under test is positioned exactly at the focal length of the beam.  In our current configuration, this distance is 15cm from the focuser, but we have also configured the system to work well for 14\,cm and 20\,cm focal lengths by simply swapping optical components.

\section{Testing Results}\label{S:results}

In testing the design described above, the primary goal was to prove the scanning functionality of the adapted MEMS mirror technology at temperatures in the milliKelvin regime. Additional goals were to demonstrate that the system is capable of achieving the design specifications of calibration beam spot size, scanning range, and repeatability.

Unless otherwise specified, the light source used in the tests described below was a ThorLabs SLS201L/M fiber-coupled, broadband tungsten-halogen lamp that emits photons over a range of 300 -- 2800\,nm.  ThorLabs NE10A and NENIR60B neutral density filters were installed at room temperature to reduce the lamp  intensity by a factor of 10 in the optical band and 500,000 in the near IR. In the DR, the OZ Optics refractive focuser upstream from the MEMS mirror was designed for a central wavelength of 650\,nm and had a focal length of 15\,cm. Lastly, a ThorLabs KG5 colored bandpass filter was inserted into the MEMS scanning unit housing output filter slot to narrow the wavelength range sent on to the device under test to 330--665\,nm. 

\subsection{Warm Characterization}

Before being operated in a dilution refrigerator, the MEMS calibration system was first tested at room temperature, where we could most easily measure the final beam spot size and intensity profile, as well as optimize the full package geometry to achieve the scanning range needed for device calibrations  Bench-top optical measurements were made using a Kralux 2.3\,megapixel CMOS camera placed at the focal point of the focuser.  For these room temperature measurements, a 0.2\,m-long custom multi-mode optical patch cable with 100\,\textmu m-core (UM22-100) was used to couple the SMA output of the light source to the FC/PC input of the focuser.

\begin{figure}[t]
    \centering
    \includegraphics[width=0.49\textwidth]{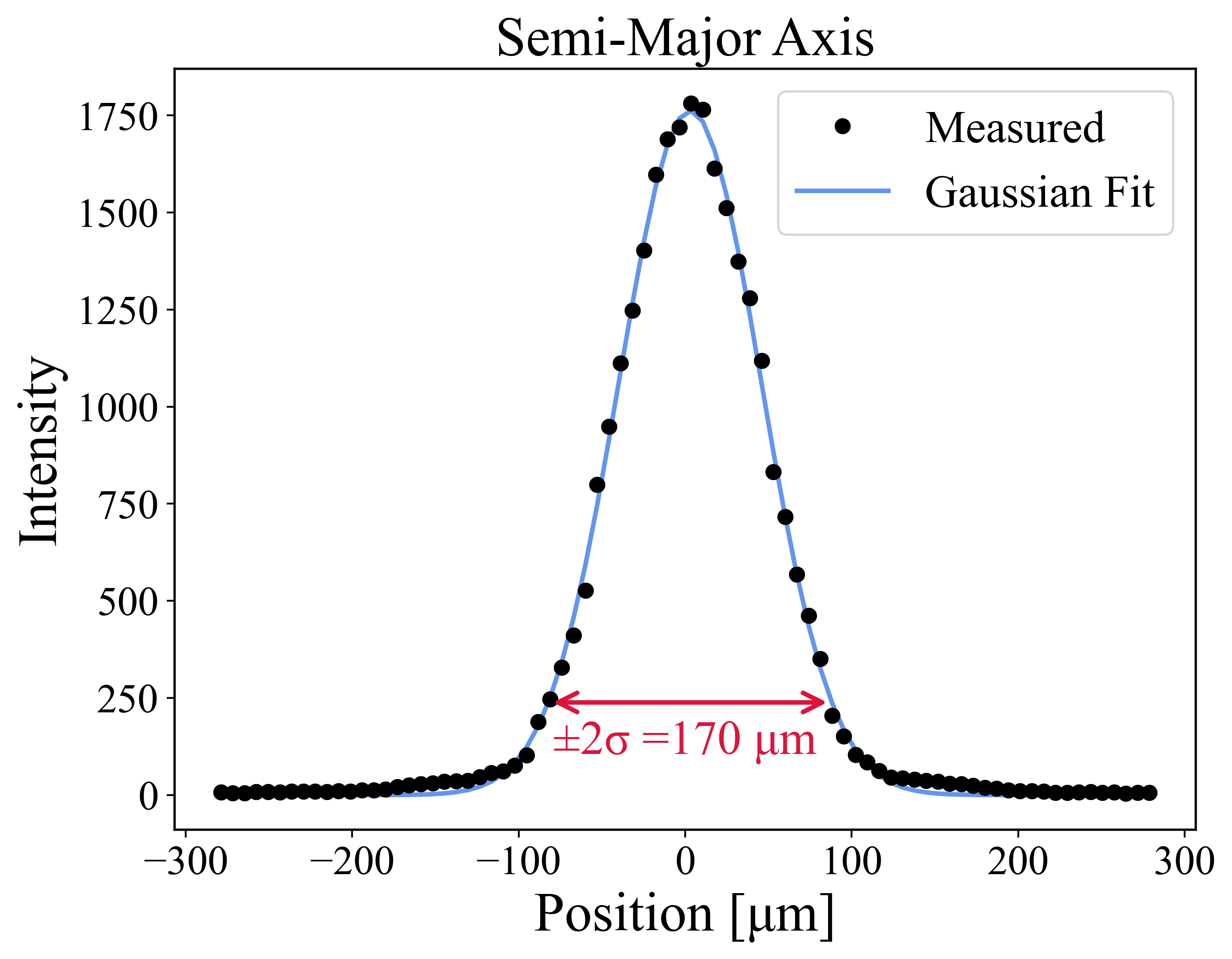}
    \includegraphics[width=0.49\textwidth]{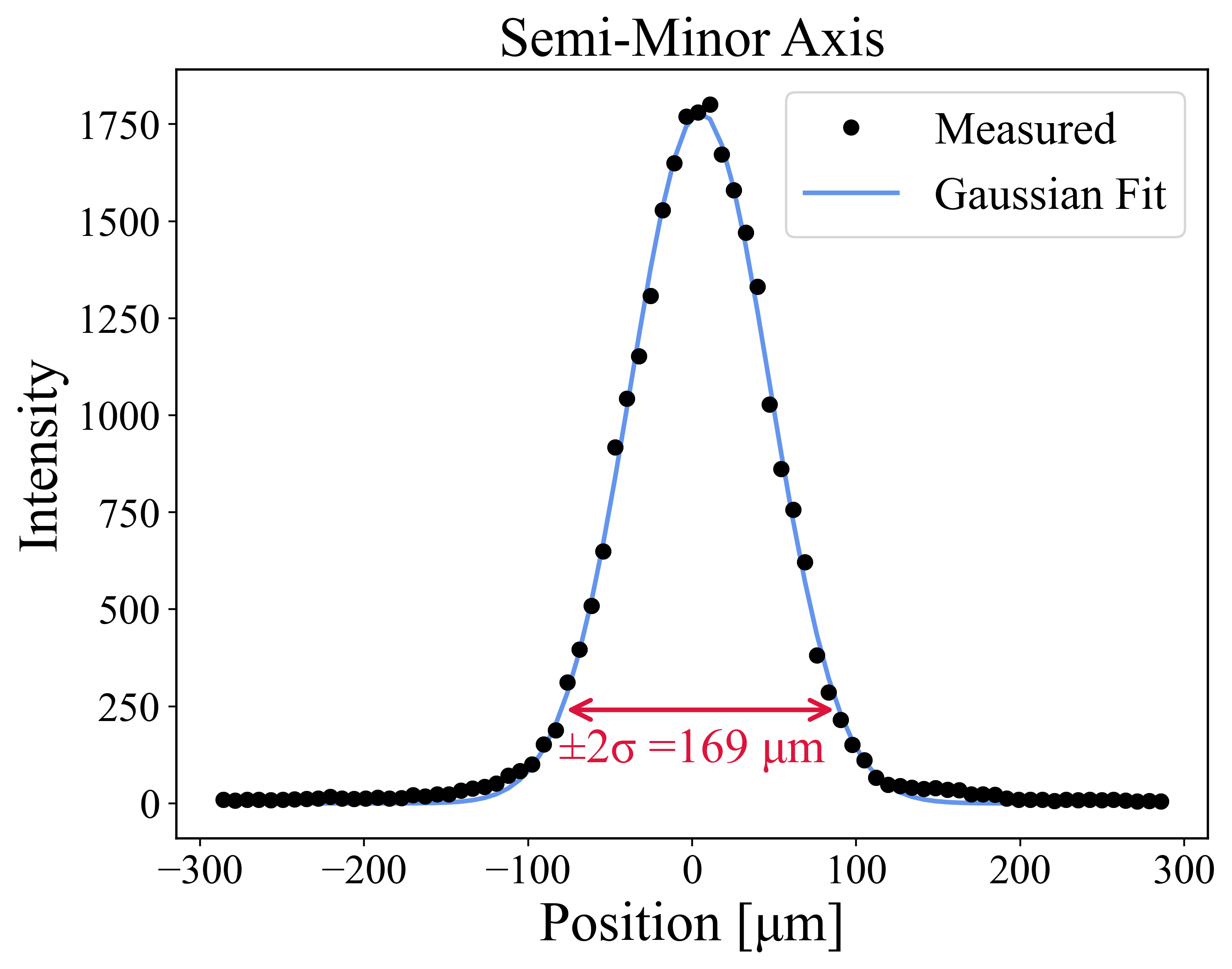}
    \caption{Distributions of pixel intensity along the semi-major (left) and semi-minor (right) axes of the beam spot. Data taken from camera images of a spot from a filtered broadband source used in conjunction with a focuser for 650\,nm light. A Gaussian fit is performed and the beam diameter is taken as the $\pm$2$\sigma$ point.~\cite{laserbeamsize}}
    \label{fig:beam_spot}
\end{figure}

Using this optical path, the beam spot diameter was measured to be $\sim$170\,\textmu m. The spot size was determined by fitting a Gaussian to the one-dimensional distributions of pixel intensity along the semi-major and semi-minor axes of the beam spot in a camera image and taking the $\pm$2$\sigma$ width~\cite{laserbeamsize},
as shown in Figure~\ref{fig:beam_spot}.  It was possible to reduce the beam spot diameter to $\sim$100\,\textmu m using additional low-pass filters that functioned to reduce chromatic aberration.  However, this additional filtering led to low intensities that were not desirable for early tests. 
In our initial bench-top tests, the MEMS mirror was directly connected to its driver via an $\sim$8"-long 10-pin plastic ribbon cable provided by Mirrorcle.

The maximum scanning area achieved was found to be $\sim$3\,cm\,$\times$\,3\,cm at the focal length of the unit, but the results varied dramatically with different cabling capacitance.  Using the cabling for which the largest scanning areas were achieved, the location of the beam spot varied linearly with applied MEMS voltage, and was reproducible upon repeated applications of the same voltages.  

\subsection{Cold Characterization}

The initial cold characterization tests of the basic setup were performed with the steering unit mounted to a device designed and fabricated by the SPT-3G+ collaboration as a prototype for the SPT-3G+ focal plane.\cite{SPTMKIDS} It was designed for measurement of the cosmic microwave background radiation in the $\sim$200-240 GHz band~\cite{SouthPoleTelescope:2021waa} and consists of an array of ten microwave kinetic inductance detectors (MKIDs). 
The MKIDs on this array are superconducting resonant circuits, consisting of an absorber, which acts as the inductive element of the circuit, coupled to an interdigitated capacitor.  In this device, both the absorber and the capacitor are fabricated with aluminum.  Optical power on the silicon substrate around each MKID generates phonons, which break Cooper pairs in the inductive element. This results in a change in the MKID's resonant frequency and quality factor that is dependent on the optical power incident on the substrate.\cite{zmuidzinas} 

\begin{figure}[t]
    \centering
    \includegraphics[width=0.95\textwidth]{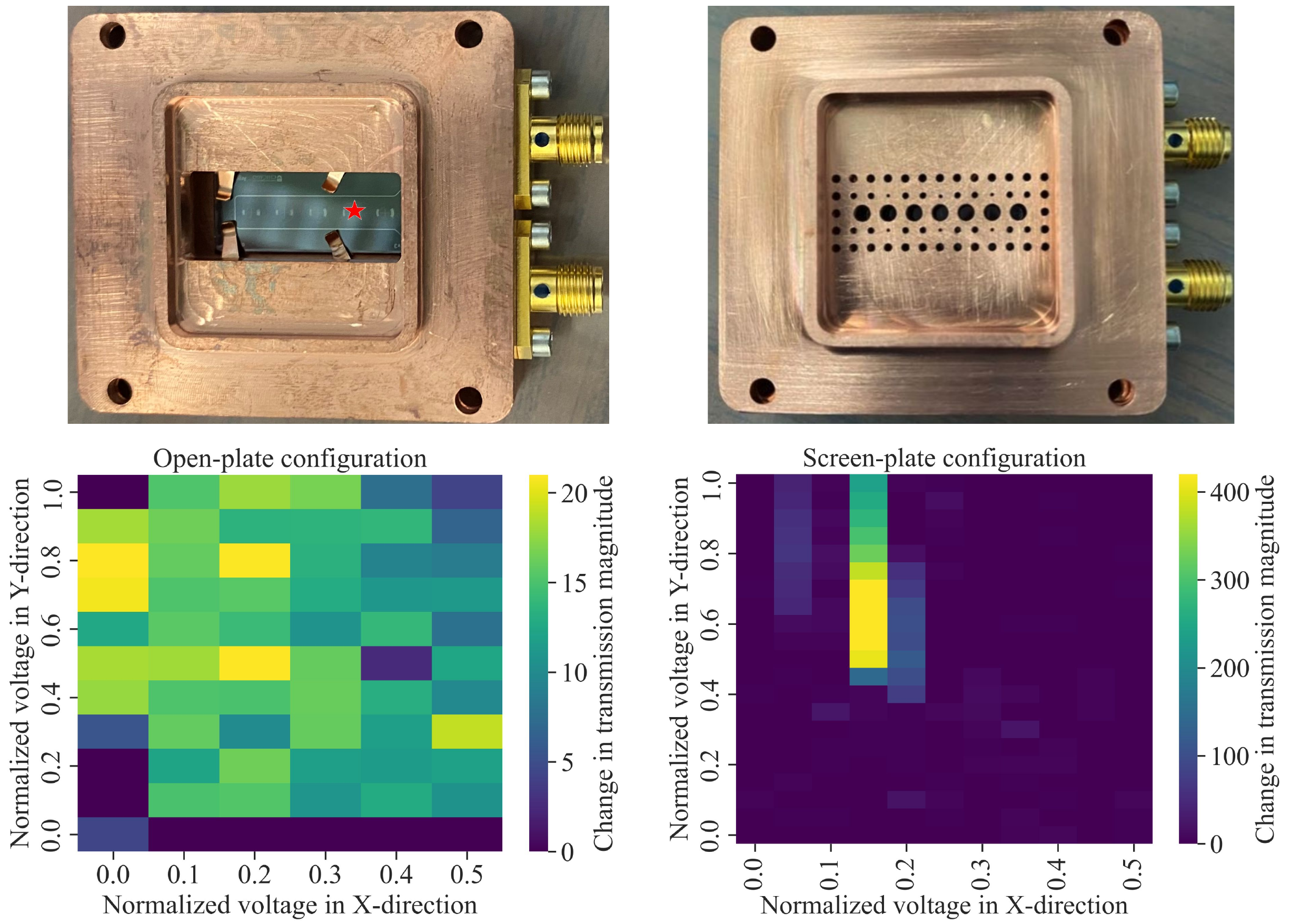}
    \caption{ 
    \textbf{Top row:} MKID detector array package with the ``open-plate'' configuration (left) and the ``screen-plate'' configuration (right). For simplicity, only one of ten MKIDs in the array was used for these studies. Its location is marked with a red star in the left image.  
    \textbf{Bottom row:} Relative response of a single MKID detector channel as a function of the MEMS mirror steering normalized control voltage in each configuration. For each data point, the MEMS mirror is steered to a location and the light source is turned on. The change in MKID transmission magnitude~($\mathbf{\Delta S_{21}}$) in response to the photon stimulation is plotted on the z-axis. MKID readout parameters and physical hardware mounting were adjusted between the two datasets, so no direct comparison regarding absolute magnitude of device response can be made between the two plots. On the left, a device response is present across much of the mapped region, except for an edge, potentially indicating a device or housing feature. On the right, substantial device response is only present in a localized region, which we correlate with the beam passing through a hole in the screen-plate. The distortion of the circular hole as measured in MEMS voltage space is attributed to non-linear mapping of voltage space to physical space, and is the subject of future improvements to the system.}
    \label{fig:scan}
\end{figure}

The MKID chip was mounted in a copper housing as shown in Figure \ref{fig:scan}, and affixed to the steering unit via the ``output plate'' shown in Figure~\ref{fig:output}, such that the beam was incident on the face of the MKID array. The scanning unit was installed and coupled to the MKID as shown in Figure~\ref{fig:output}. Data was taken in two geometric configurations. In the first configuration, a lid was attached to the MKID housing with a single large opening that exposed the entire 10-MKID array to incident photons (``open-plate configuration").  A second run was performed later using a machined copper plate containing a known hole pattern (``screen-plate configuration").

Once the detector and MEMS scanning package was fully assembled, the combined unit was mounted directly to the mixing chamber plate of a Bluefors LD-400 with a nominal base temperature of $<$10mK and 600uW of cooling power at 100mK. 
For the duration of these tests, the base temperature of the dilution unit was maintained between 15 and 35\,mK.  
The MKID was read out using the QICK instrumentation control package~\cite{QICK2022,Smith:2022goi}.  The QICK system consisted of a Xilinx ZCU111 evaluation board with a XCZU28DR RFSoC FPGA. The ZCU111 ran custom firmware based on an overlapped polyphase filterbank~\cite{price2018spectrometers}, which channelizes data into 2 MHz bandwidth per tone and was capable of streaming 1024 channels per RF line. The interface for controlling the firmware and streaming data was provided through a Jupyter notebook implemented through the PYNQ environment~\cite{PYNQ}.

Prior to taking data with the MEMS scanning system, a frequency scan was performed for all the MKIDs on the array to ensure that they were operational and to evaluate their basic quality.  The MKID with the highest Q (measured to be $\sim$ $1.2 \times 10^5$) was used for two initial characterization studies performed with the ``open-plate" and ``screen-plate" configurations respectively.  The high-Q pixel on the MKID chip is oriented vertically at the physical location marked with a star in Figure~\ref{fig:scan}.

\begin{figure}[t]
    \centering
    \includegraphics[width=0.7\textwidth]{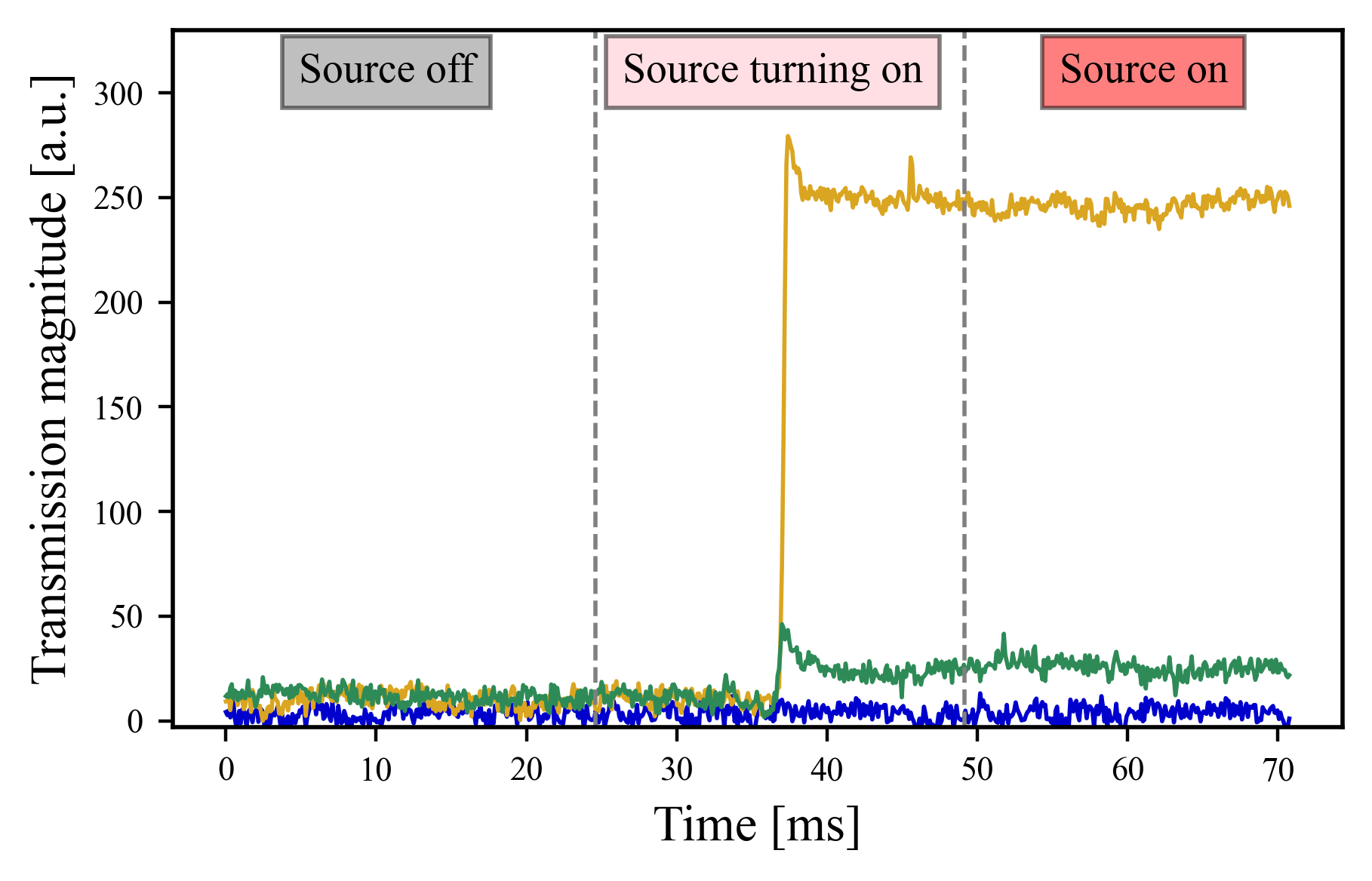}
    \caption{
        Time-domain MKID response in the ``open-plate'' configuration as the light source is turned on. Three traces show that response varies dramatically when the beam is directed to different locations across the MKID array.  
        Yellow ([0.9, 0.65] in MEMS voltage space) shows the highest response, green (0.3,0.3) shows a modest response, and blue (0.25,0.15) shows negligible response.}
    \label{fig:laseronoff}
\end{figure}

The first cryogenic run (``open-plate configuration") successfully demonstrated movement of the MEMS mirror at 25\,mK. The scanning test consisted of stepping the MEMS mirror position through the two-dimensional voltage space of its controller. For each position in voltage space, RF transmission measurements were made at the MKID resonance frequency both before and after the continuous-wave light source was switched on.  In the presence of incident photons, we expect the MKID's resonance to shift downwards in frequency and decrease in quality factor~\cite{temples2024performance}.  Both of these effects will lead to an increase in transmission magnitude as measured at a fixed readout frequency located on the original resonance feature. An example of this photon-induced transmission change is shown in Figure~\ref{fig:laseronoff}. From these measurements, a differential in the MKID transmission magnitude resulting from the light source was then constructed.  Following each measurement, a wait time of approximately 20 seconds was observed to allow the chip to cool and the resonance to return to its original location in frequency space.

To map out the response across the plane of the sensor, a thorough scan through a section of the MEMS voltage space was performed following the data-taking procedure described above.  The mirror was stepped through linear increments of the voltage in both X and Y, thus creating a grid of MKID response ``pixels".  This two-dimensional response map from the open-plate configuration run is shown in the left-hand side of Figure~\ref{fig:scan}.  The data shows clear regions of strong response to the photon beam (light colored pixels).  These are delineated from regions of weak response (dark colored pixels).

The first dataset acquired while running with the open-plate configuration is difficult to fully interpret. We saw varying but repeatable detector response as a function of mirror position that was suggestive of beam steering, but it was not immediately obvious which location along the MKID chip was being targeted by the beam. We hypothesized that this was due to uncontrolled reflections creating a source of background noise seen by the MKID array. This hypothesis was supported by warm testing results where we intentionally used the MEMS mirror to steer the beam off-chip and saw large streaks of light across the device housing when the beam was reflected off areas of the output plate of the copper enclosure.

To mitigate these reflections and aid in subsequent data analysis, we introduced a machined copper “screen-plate” to our detector mounting set-up, as shown in the upper right of Figure~\ref{fig:scan}. This plate is mounted in front of the MKID chip as shown in Fig.~\ref{fig:output}, such that large regions of the MKID chip are masked. The same MKID device was tested again with much cleaner results, as shown in the lower right of Figure~\ref{fig:scan}. In particular, the single instrumented MKID on the detector chip clearly responded when exposed to steered photons, as evidenced by region of high response when the beam was steered over a hole in the screen-plate. While these holes were circular, we observed an elongated density profile in the data, implying a nonlinear mapping from MEMS voltage coordinates to physical space. 

We were able to show that this image distortion was caused by our cabling. In particular, when using higher capacitance cables, MEMS mirror movement is reduced near the edges of the scanning region relative to the central region. The result is that voltage changes made near the center of the scan region correspond to a larger step in physical position of the beam than the same voltage change applied near the edge of the scanning range. This model accounts well for the signal pattern elongation shown in the lower right of Figure~\ref{fig:scan}. Further work to mitigate these unwanted distortions is ongoing and will be discussed in Section~\ref{S:discussion}.

\begin{figure}[t]
    \centering
    \includegraphics[width=0.65\textwidth]{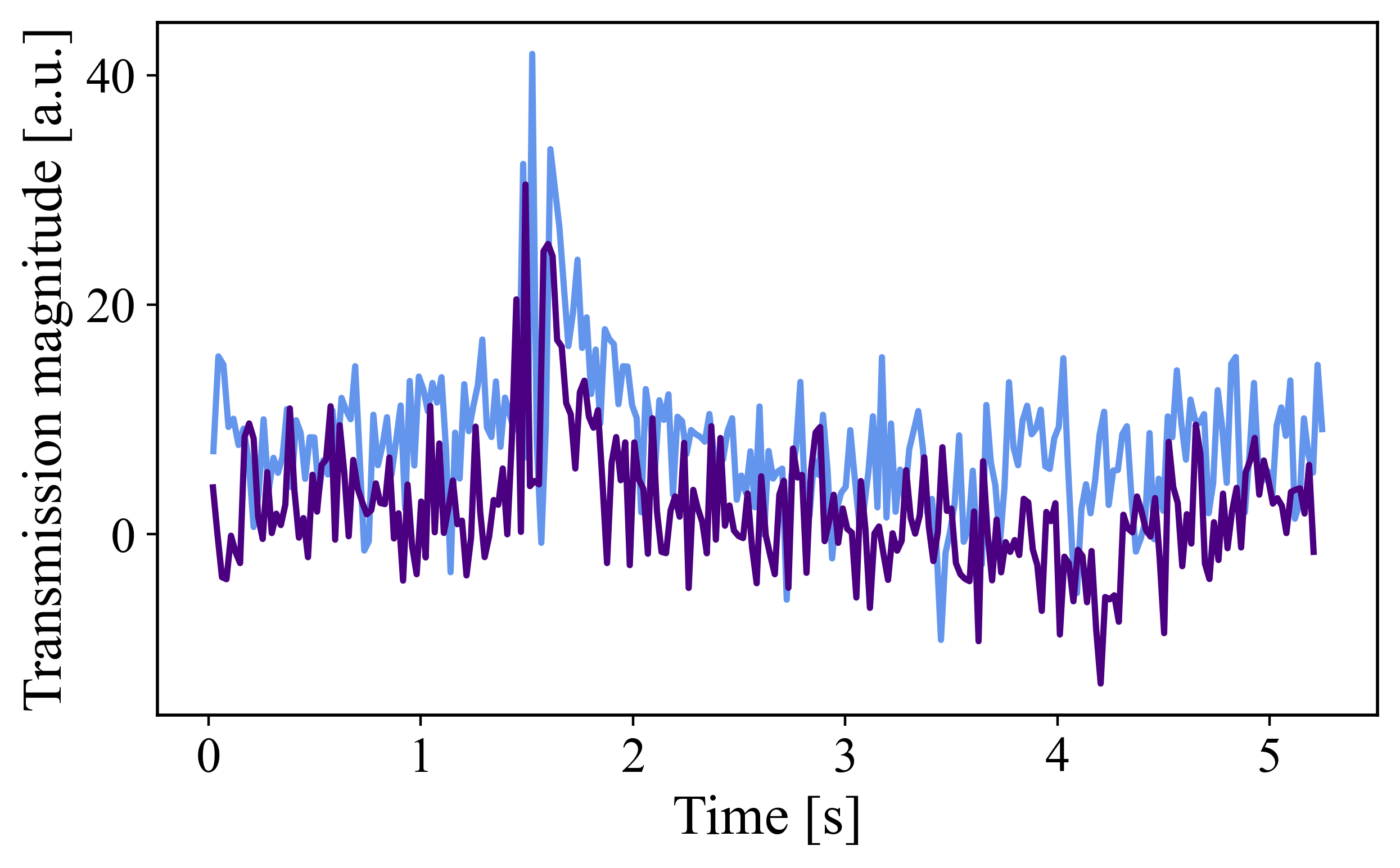}
    \caption{Two time traces, acquired $\sim$60~minutes apart, showing the MKID response as the  MEMS mirror was scanned diagonally in voltage space coordinates from [0.4, 0.4] to [0.1, 0.1]. The observed change in MKID response occurred when the mirror was positioned at [0.3, 0.3] in voltage space, corresponding to the steered photon beam passing over a hole in the screen-plate. The similarity of the two data sets (dark blue and light blue) suggest a high degree of stability and repeatability in the full MEMS-based calibration set-up. These data were taken with the same hardware configuration and readout parameters as the masked “screen-plate” data shown in Figure~\ref{fig:scan}.}
    \label{fig:1Dscanning}
\end{figure}

The consistency of the MEMS beam-positioning capability at $\sim$20\,mK was studied by scanning the mirror along a diagonal path in voltage space and monitoring the resulting MKID signal. This scan was repeated multiple times over the course of an hour. Typical results are shown in Figure~\ref{fig:1Dscanning}, where a similar change in MKID signal is seen to occur at the same location for two independent mirror scans. This clearly demonstrates active and repeatable cryogenic beam steering from the MEMS mirror. 

To test the power dissipation of the mirror, a heater on the mixing chamber of the dilution fridge was assigned a temperature setpoint above the base temperature of the fridge.  Once equilibrium was established, the MEMS mirror was biased and set to trace across its maximum range at its highest allowable speed.  The power of the fridge heater was monitored, and was found to remain within its level of noise within the duration of the test.  In a different MEMS unit tested in a separate dilution fridge facility, an analysis of thermometry data demonstrated that the MEMS unit dissipated an average of 0.99$\,\pm\,$0.28 \textmu W of power during a resting power-on state, with minimal additional dissipation during periodic mirror rotations occurring at a rate of 1\,Hz.

\section{Discussion and Follow-up Work}\label{S:discussion}

The initial cold characterization of the prototype MEMS steering unit clearly demonstrated the successful operation of the MEMS mirror at sub-Kelvin temperatures. Still, drawbacks to the design and setup were discovered during the first round of testing. For this reason, several follow-up measurements were performed and many design upgrades are planned for future iterations of the calibration system.

\subsection{Focusing unit}

While a beam spot diameter of $\sim$200\,\textmu m was achieved in these tests, the specifications for the focuser give an idealized value of $\sim$80\,\textmu m. This difference is likely due to the mismatch between the relatively wide spectrum of the filtered broadband light source used for these tests and the narrow band of wavelengths for which the focuser is designed. To explore whether an improved beam spot diameter is achievable in our system, a follow-up measurement was performed using a 625\,nm fiber-coupled LED and a OZ Optics focuser designed for 635\,nm light with a 15\,cm focal length. In this warm test, a minimum beam spot diameter of 80\,\textmu m was achieved.

Still, this focusing method only works for the single wavelength matched to the focuser design, and the focuser design is restricted to the wavelengths between 180 -- 2000\,nm, which is the current limiting factor for the energy range of the overall calibration system. If subsequent calibrations want to use a multiple wavelengths over a wider range, the dilution refrigerator would have to be warmed up, the calibration system uninstalled, the focuser replaced, and the calibration system reinstalled. For this reason, the commercial refractive focusing unit will be replaced in future iterations of the calibration system with a custom reflective focusing system that provides comparable or even improved beam spot size diameters and works over a wide range of wavelengths without needing to be replaced.

\subsection{Electrical cabling}

During characterization following installation into the dilution refrigerator it was noted that the beam spot position no longer moved linearly with the applied voltage and the scanning area was significantly reduced as compared to initial warm bench tests. Additionally, on three separate occurrences during installation and validation testing, unstable behavior was observed for specific combinations of bias voltage. In this coordinate of voltage space, the mirror was unable to hold a steady position, and the beam spot was observed to oscillate rapidly across a linear path with length on the order of 500\,\textmu m. Note that the location of this instability was excluded from the data-taking process that was used to create Figure~\ref{fig:scan}. 

These behaviors were observable only when using the electrical cabling required for operation in the dilution refrigerator. In followup investigations, it was discovered that this was due to the intermediate air-side cable used to carry the DC voltages to the MEMS mirror, which was a set of twelve, individually shielded twisted pairs terminating in a Fischer connector. In its current configuration, the commercial DC bias controller is capable of driving up to 50\,pF per channel. Since the capacitance of the MEMS mirror itself is approximately 20\,pF, this leaves a maximum capacitance in the cabling of about 30\,pF before there are adverse effects. We observed that exceeding this limit led to the observed reduced steering range and non-linear position response to applied voltage near the edges of the range.

For future experiments, we have developed a simplified electrical path that removes the need for intermediate air-side cabling and reduces the number of connections inside the dilution refrigerator itself. The MEMS mirror still connects to the 10-pin/MDM adapter board embedded in one wall of the MEMS scanning unit enclosure via a short, plastic 10-pin ribbon cable, but the adapter board is connected all the way to a DB25 vacuum feedthrough at the 300K stage via a 2\,m-long Constantan woven loom with twelve twisted pairs that is thermally fixed at each stage of the dilution refrigerator. The air side of the DB25 feedthrough now connects directly to the DB25/10-pin adapter board, which is connected to the MEMS driver via short, plastic 10-pin ribbon cable. Warm tests utilizing this setup showed the full expected scanning range of $\sim$3\,cm$\times$\,3\,cm and improved linearity of beam spot location with applied voltage. Final cold tests of the system are underway.

\subsection{Other improvements}

Another challenge to address is the possible reflections of the beam off of bare copper surfaces in the MEMS steering unit enclosure. To limit this effect, as well as any other source of stray photons in the MEMS steering unit housing that may diminish cryogenic device performance, we plan to coat all inner surfaces with a cryogenic- and vacuum-compatible black paint~\cite{aeroglaze} or an epoxy mixed with carbon lumped black~\cite{BockThesis}.

Finally, the overall footprint of the this prototype calibration system in a dilution refrigerator is still relatively large. This leads to inefficient usage of space, and also can make it challenging to couple the calibration system to magnetic shielding layers that are required for various sensor types. For this reason, a second-generation design has been produced. The new design has a smaller and simplified layout, which allows the entire system to fit within A4K shielding cans, and also reduces the cost.

\section{Conclusion}\label{S:conclusion}

There is a distinct need in both the particle astrophysics and quantum information science communities to characterize sensors in cryogenic environments. 
Current techniques typically focus on energies above the range of interest for these devices, are limited by a lack of position control, offer limited coverage area, struggle with position resolution due to large spot size or hysteresis, or cannot operate at cryogenic temperatures due to heat dissipation or freeze-out of control mechanisms. 
The MEMS mirror-based calibration system presented here overcomes many of these limitations. 

Tests of our prototype MEMS scanning unit successfully demonstrated a $\sim$100\,\textmu m beam spot that can be steered over an area of $\sim$3\,cm$\times$\,3\,cm down to 20\,mK with $<$1\,\textmu W of power dissipation, making this the first sub-Kelvin operation of the MEMS-mirror technology. The optical system used in these tests can deliver of photons over an energy range of 0.6 -- 5\,eV. Despite some issues with high-capacitance electrical cabling, the positioning of the beam was found to be reproducible, achieving the goal of repeated photon injections on the surface of a device through cryogenic beam steering. Given several small drawbacks of the original design, a next-generation version with an improved electrical path, optical path, and mechanical design is in development.

We expect the new iteration of our system to be useful soon for a range of exciting applications, including the characterization of background events in dark matter detectors and investigation of underlying physical mechanisms driving superconducting qubit decoherence. 
\section{Acknowledgements}

The authors acknowledge Robert Moffatt and Blas Cabrera as the architects of the first MEMS-mirror-based beam steering effort, on which this system is based~\cite{Moffatt:2016kok}.  We thank the QSC Postdoctoral Research Awards for providing the seed money that kick-started this project.  We thank Mirrorcle Technologies, Inc. for working with us to produce a version of their MEMS mirror that functions at cryogenic temperatures, and the DMQIS group at SLAC for their frequent help with lab space and other logistics. HM would like to thank Joseph Angelo and Kavin Ammigan for mentorship and CAD advice as part of the Fermilab Engineering Co-op Program, as well as Tyler Funk for discussions surrounding cryogenic design. KS would like to thank Alden Fan for his help with beam spot size analysis and Aviv Simchony for his help with breakout board construction.

This manuscript has been authored by Fermi Research Alliance, LLC under Contract No. DE-AC02-07CH11359 with the U.S. Department of Energy, Office of Science, Office of High Energy Physics. This work was supported by the U.S. Department of Energy, Office of Science, National Quantum Information Science Research Centers, Quantum Science Center and the U.S. Department of Energy, Office of Science, High Energy Physics Program Office. This work was supported in part by the U.S. Department of Energy, Office of Science, Office of Workforce Development for Teachers and Scientists (WDTS) under the Science Undergraduate Laboratory Internships Program (SULI). This work was also made possible in part through a KIPAC Innovation Grant and support from the KIPAC Post-baccalaureate Fellowship Program. 

This project was originally conceived and developed by NK, and was guided to completion by KS. 
HM, KS, and NK designed the physical setup of the scanning unit, including the housing body, electrical path, and optical path. 
IH performed early warm characterization of the MEMS mirrors. 
HM and KS performed warm validations of the entire system prior to installation and were also responsible for installation. 
AA provided the MKID device, which was designed and fabricated by KD and PB. 
AA and MY provided space in their dilution refrigerator and managed cooldowns. 
HM, AN, KA, and KS performed the tests to determine MEMS power dissipation, after which KA and RL analyzed the resulting data. 
HM, AA, DT, and KS set up the MKID readout electronics using a QICK RFSoC package developed by CS, SU, GC, and LS. 
SU additionally provided invaluable debugging during early data-taking. 
HM and KS collected the MKID data presented in the paper, and AA was instrumental in helping them understand it. 
HM was responsible for the final analysis of the MKID data and for making all plots. 
KS, HM, LH, BY, NT, SM, and DBa were the primary authors of this report. 
DBa, DBo, RK, LH, NK, and DT provided valuable feedback on the project at weekly meetings. 
RL provided many useful conversations to help us understand various signals in the data. 
DBa, DBo, AC, EFF, LH, RK, NK, and BY provided research group leadership for the duration of the project. 
\bibliography{references}{}

\begin{thebibliography}{10}

\bibitem{Kolb:1990vq}
E.W. Kolb and M.S. Turner.
\newblock {\em {The Early Universe}}, volume~69.
\newblock 1990.

\bibitem{2020A&A...641A...6P}
N.~{Aghanim} et~al.
\newblock {Planck 2018 results. VI. Cosmological parameters}.
\newblock {\em Astronomy \& Astrophysics}, 641:A6, September 2020.

\bibitem{Cooley:2022ufh}
J.~Cooley et~al.
\newblock {Report of the Topical Group on Particle Dark Matter for Snowmass 2021}.
\newblock 9 2022.

\bibitem{LZ}
J.~Aalbers et~al.
\newblock {First Dark Matter Search Results from the LUX-ZEPLIN (LZ) Experiment}.
\newblock {\em Phys. Rev. Lett.}, 131:041002, Jul 2023.

\bibitem{PICO:2019vsc}
C.~Amole et~al.
\newblock {Dark Matter Search Results from the Complete Exposure of the PICO-60 C$_3$F$_8$ Bubble Chamber}.
\newblock {\em Phys. Rev. D}, 100(2):022001, 2019.

\bibitem{PDG}
R.L. Workman et~al.
\newblock {Review of Particle Physics}.
\newblock {\em PTEP}, 2022:083C01, 2022.

\bibitem{SnowmassLowThresholdReport}
R.~{Essig} et~al.
\newblock {Snowmass2021 Cosmic Frontier: The landscape of low-threshold dark matter direct detection in the next decade}.
\newblock {\em arXiv e-prints}, page arXiv:2203.08297, March 2022.

\bibitem{SENSEI}
P.~Adari et~al.
\newblock {SENSEI: First Direct-Detection Results on sub-GeV Dark Matter from SENSEI at SNOLAB}.
\newblock 12 2023.

\bibitem{DAMIC-M}
I.~Arnquist et~al.
\newblock {First Constraints from DAMIC-M on Sub-GeV Dark-Matter Particles Interacting with Electrons}.
\newblock {\em Phys. Rev. Lett.}, 130(17):171003, 2023.

\bibitem{Ren:2020gaq}
R.~Ren et~al.
\newblock {Design and characterization of a phonon-mediated cryogenic particle detector with an eV-scale threshold and 100~keV-scale dynamic range}.
\newblock {\em Phys. Rev. D}, 104(3):032010, 2021.

\bibitem{temples2024performance}
D.J. Temples et~al.
\newblock Performance of a kinetic inductance phonon-mediated detector at the nexus cryogenic facility, 2024.

\bibitem{Cardani:2021iff}
L.~Cardani, N.~Casali, I.~Colantoni, A.~Cruciani, S.~Di~Domizio, M.~Martinez, V.~Pettinacci, G.~Pettinari, and M.~Vignati.
\newblock {Final results of CALDER: kinetic inductance light detectors to search for rare events}.
\newblock {\em Eur. Phys. J. C}, 81(7):636, 2021.

\bibitem{Cruciani:2022mbb}
A.~Cruciani et~al.
\newblock {BULLKID: Monolithic array of particle absorbers sensed by kinetic inductance detectors}.
\newblock {\em Appl. Phys. Lett.}, 121(21):213504, 2022.

\bibitem{Luskin:2023ksc}
J.S. Luskin et~al.
\newblock {Large active-area superconducting microwire detector array with single-photon sensitivity in the near-infrared}.
\newblock {\em Appl. Phys. Lett.}, 122(24):243506, 2023.

\bibitem{linehan2024estimating}
R.~Linehan, I.~Hernandez, D.~J. Temples, S.~Q. Dang, D.~Baxter, L.~Hsu, E.~Figueroa-Feliciano, R.~Khatiwada, K.~Anyang, D.~Bowring, G.~Bratrud, G.~Cancelo, A.~Chou, R.~Gualtieri, K.~Stifter, and S.~Sussman.
\newblock Estimating the energy threshold of phonon-mediated superconducting qubit detectors operated in an energy-relaxation sensing scheme, 2024.

\bibitem{Fink:2023tvb}
C.W. Fink, C.P. Salemi, B.A. Young, D.I. Schuster, and N.A. Kurinsky.
\newblock {The Superconducting Quasiparticle-Amplifying Transmon: A Qubit-Based Sensor for meV Scale Phonons and Single THz Photons}.
\newblock 10 2023.

\bibitem{Wilen}
C.~D. Wilen, S.~Abdullah, N.~A. Kurinsky, C.~Stanford, L.~Cardani, G.~D’Imperio, C.~Tomei, L.~Faoro, L.~B. Ioffe, C.~H. Liu, A.~Opremcak, B.~G. Christensen, J.~L. DuBois, and R.~McDermott.
\newblock Correlated charge noise and relaxation errors in superconducting qubits.
\newblock {\em Nature}, 594(7863):369–373, June 2021.

\bibitem{Harrington:2024iqm}
P.M. Harrington et~al.
\newblock {Synchronous Detection of Cosmic Rays and Correlated Errors in Superconducting Qubit Arrays}.
\newblock 2 2024.

\bibitem{Vepsalainen:2020trd}
A.~Veps\"al\"ainen et~al.
\newblock {Impact of ionizing radiation on superconducting qubit coherence}.
\newblock {\em Nature}, 584(7822):551--556, 2020.

\bibitem{McEwen}
M.~McEwen et~al.
\newblock Resolving catastrophic error bursts from cosmic rays in large arrays of superconducting qubits.
\newblock {\em Nat. Phys.}, 18:107--111, 2022.

\bibitem{McEwen:2024nrv}
M.~McEwen et~al.
\newblock {Resisting high-energy impact events through gap engineering in superconducting qubit arrays}.
\newblock 2 2024.

\bibitem{Iaia_2022}
V.~Iaia, J.~Ku, A.~Ballard, C.~P. Larson, E.~Yelton, C.~H. Liu, S.~Patel, R.~McDermott, and B.~L.~T. Plourde.
\newblock Phonon downconversion to suppress correlated errors in superconducting qubits.
\newblock {\em Nature Communications}, 13(1), October 2022.

\bibitem{Martinis:2020fxb}
J.M. Martinis.
\newblock {Saving superconducting quantum processors from decay and correlated errors generated by gamma and cosmic rays}.
\newblock {\em npj Quantum Inf.}, 7:90, 2021.

\bibitem{Baxter:2022dkm}
D.~Baxter et~al.
\newblock {Snowmass2021 Cosmic Frontier White Paper: Calibrations and backgrounds for dark matter direct detection}.
\newblock 3 2022.

\bibitem{10.1063/5.0022533}
M.~Hegedüs, K.~Fedorov, I.~Antonov, P.~Karataev, and V.~N. Antonov.
\newblock {Detection of black body radiation using a compact terahertz imager}.
\newblock {\em Applied Physics Letters}, 117(23):231106, 12 2020.

\bibitem{Moffatt:2016kok}
R.~Moffatt.
\newblock {\em {Two-Dimensional Spatial Imaging of Charge Transport in Germanium Crystals at Cryogenic Temperatures}}.
\newblock PhD thesis, Stanford U., 2016.

\bibitem{10.1063/1.4939753}
R.~A. Moffatt, B.~Cabrera, B.~M. Corcoran, J.~M. Kreikebaum, P.~Redl, B.~Shank, J.~J. Yen, B.~A. Young, P.~L. Brink, M.~Cherry, A.~Tomada, A.~Phipps, B.~Sadoulet, and K.~M. Sundqvist.
\newblock {Imaging the oblique propagation of electrons in germanium crystals at low temperature and low electric field}.
\newblock {\em Applied Physics Letters}, 108(2):022104, 01 2016.

\bibitem{10.1063/1.5131171}
C.~Stanford, R.~A. Moffatt, N.~A. Kurinsky, P.~L. Brink, B.~Cabrera, M.~Cherry, F.~Insulla, M.~Kelsey, F.~Ponce, K.~Sundqvist, S.~Yellin, and B.~A. Young.
\newblock {High-field spatial imaging of charge transport in silicon at low temperature}.
\newblock {\em AIP Advances}, 10(2):025316, 02 2020.

\bibitem{10.1063/5.0038392}
C.~Stanford, M.~J. Wilson, B.~Cabrera, M.~Diamond, N.~A. Kurinsky, R.~A. Moffatt, F.~Ponce, B.~von Krosigk, and B.~A. Young.
\newblock {Photoelectric absorption cross section of silicon near the bandgap from room temperature to sub-Kelvin temperature}.
\newblock {\em AIP Advances}, 11(2):025120, 02 2021.

\bibitem{2023APS..APRT12005O}
D.~{Osterman}, D.~{Jin}, X.~{Han}, X.~{Li}, X.~{Zhou}, Y.~{Huang}, and {Spice/Herald Collaboration Team}.
\newblock {Measuring Quasiparticle Diffusion in Superconducting Aluminum Films with a TES and Microscopic Laser-Scanning Technique}.
\newblock In {\em APS April Meeting Abstracts}, volume 2023 of {\em APS Meeting Abstracts}, page T12.005, January 2023.

\bibitem{PhysRevLett.119.223602}
D.~D. Sukachev, A.~Sipahigil, C.~T. Nguyen, M.~K. Bhaskar, R.~E. Evans, F.~Jelezko, and M.~D. Lukin.
\newblock Silicon-vacancy spin qubit in diamond: A quantum memory exceeding 10 ms with single-shot state readout.
\newblock {\em Phys. Rev. Lett.}, 119:223602, Nov 2017.

\bibitem{10.1117/12.2595780}
B.~J. Lawrie, M.~Feldman, C.~E. Marvinney, and Y.~Y. Pai.
\newblock {Free-space confocal magneto-optical spectroscopies at milliKelvin temperatures}.
\newblock In Cesare Soci, Matthew~T. Sheldon, and Mario Agio, editors, {\em Quantum Nanophotonic Materials, Devices, and Systems 2021}, volume 11806, page 1180604. International Society for Optics and Photonics, SPIE, 2021.

\bibitem{Benevides:2023ldf}
R.~Benevides, M.~Drimmer, G.~Bisson, F.~Adinolfi, U.~von L\"upke, H.M. Doeleman, G.~Catelani, and Y~Chu.
\newblock {Quasiparticle dynamics in a superconducting qubit irradiated by a localized infrared source}.
\newblock 12 2023.

\bibitem{naftaly10.1364/AO.50.003201}
M.~Naftaly and R.~Dudley.
\newblock Terahertz reflectivities of metal-coated mirrors.
\newblock {\em Applied optics}, 50:3201--4, 07 2011.

\bibitem{laserbeamsize}
S.~Prahl.
\newblock laserbeamsize python package, 2023.

\bibitem{SPTMKIDS}
K.~Dibert, P.~Barry, Z.~Pan, A.~Anderson, B.~Benson, C.~Chang, K.~Karkare, J.~Li, T.~Natoli, M.~Rouble, E.~Shirokoff, and A.~Stark.
\newblock Development of mkids for measurement of the cosmic microwave background with the south pole telescope.
\newblock {\em Journal of Low Temperature Physics}, 209(3–4):363–371, June 2022.

\bibitem{SouthPoleTelescope:2021waa}
K.~Dibert et~al.
\newblock {Development of MKIDs for Measurement of the Cosmic Microwave Background with the South Pole Telescope}.
\newblock {\em J. Low Temp. Phys.}, 209(3-4):363--371, 2022.

\bibitem{zmuidzinas}
J.~Zmuidzinas.
\newblock Superconducting microresonators: Physics and applications.
\newblock {\em Annual Review of Condensed Matter Physics}, 3(Volume 3, 2012):169--214, 2012.

\bibitem{QICK2022}
L.~Stefanazzi et~al.
\newblock {The QICK (Quantum Instrumentation Control Kit): Readout and control for qubits and detectors}.
\newblock {\em Rev. Sci. Instrum.}, 93(4):044709, 2022.

\bibitem{Smith:2022goi}
J.P. Smith, J.I. {Bailey III}, and B.A. Mazin.
\newblock {Highly-Multiplexed Superconducting Detector Readout: Approachable High-Speed FPGA Design}.
\newblock In {\em {30th IEEE International Symposium on Field-Programmable Custom Computing Machines}}, 3 2022.

\bibitem{price2018spectrometers}
D.C. Price.
\newblock Spectrometers and polyphase filterbanks in radio astronomy, 2018.

\bibitem{PYNQ}
Python productivity for zynq.
\newblock \url{http://www.pynq.io}.

\bibitem{aeroglaze}
D.M. McCroskey, G.C. Abell, and M.H. Chidester.
\newblock {Aeroglaze Z306 black paint for cryogenic telescope use: outgassing and water vapor regain}.
\newblock In Philip T.~C. Chen and O.~Manuel Uy, editors, {\em Optical Systems Contamination and Degradation II: Effects, Measurements, and Control}, volume 4096, pages 119 -- 128. International Society for Optics and Photonics, SPIE, 2000.

\bibitem{BockThesis}
J.J. Bock.
\newblock {\em Rocket-Borne Observation of Singly Ionized Carbon 158um Emission from the Diffuse Interstellar Medium}.
\newblock PhD thesis, University of California, Berkeley, 1994.

\end{thebibliography}

\end{spacing}
\end{document}